\documentstyle[prl,aps,multicol]{revtex}
\input epsf
\begin{document}
\draft
\widetext
\title{Search for incoherent tunnel fluctuations of the magnetisation in
nanoparticles of artificial ferritin.} 
\author{C. Gilles, P. Bonville}
\address{CEA, C.E. Saclay, Service de Physique de l'Etat Condens\'e, 91191 Gif
-sur-Yvette, France}
\author{K. K. W. Wong, S. Mann}
\address{School of Chemistry, University of Bristol, England}
\date{\today}\maketitle\widetext
\begin{abstract}
\leftskip 54.8pt
\rightskip 54.8pt
The magnetic behaviour of nanoparticles of antiferromagnetic artificial
ferritin, with a mean Fe loading of 410 atoms per core, has been investigated 
by $^{57}$Fe M\"ossbauer absorption spectroscopy
down to very low temperature (34\,mK).
In previous experiments of frequency-dependent magnetic susceptibility 
$\chi(\omega)$ and magnetic noise $S(\omega$) performed at 25\,mK in similar 
samples, it was claimed that a resonance at a frequency of about 10$^8$\,Hz, 
due to a macroscopic quantum coherent state, had been observed.
However, our search of incoherent tunnel fluctuations
around 10$^8$\,Hz using $^{57}$Fe M\"ossbauer spectroscopy, whose ``window'' of
measurement of fluctuation frequencies lies in this frequency range, was
unsuccessful. This casts a doubt about the previous observation of macroscopic
quantum coherence in ferritin.
\end{abstract} 
\begin{multicols}{2}
\narrowtext

It was recently shown \cite{hemmen} that in small ferromagnetic (FM)
particles, a macroscopically large number of spins coupled by a strong 
exchange interaction can coherently tunnel through 
the energy barrier created by magnetic anisotropy \cite{gunther}.
Similar, and even stronger, quantum effects should exist in 
the antiferromagnetic (AF) particles where tunneling of the N\'eel vector 
$\bf l$ must result in a quantum superposition of AF sublattices 
\cite{barbara,chudnovsky,chud2}. 
Two different situations must be distinguished: on one side, macroscopic 
quantum tunneling (MQT) where the magnetisation incoherently
tunnels through the energy barrier, and on the other side, macroscopic 
quantum coherence (MQC) where the magnetisation coherently oscillates between 
the classically degenerate directions.
Because of the AF ordering and the small size of the cores, 
ferritin has been thought to be a good candidate for the observation of these
quantum effects \cite{barbara,chudnovsky,chud2}.
The observation of a low temperature resonance around 10$^8$\,Hz
in the absorption spectrum of artificial ferritin with an Fe loading of 1000
atoms per core \cite{awschalom,gider} was interpreted
as due to the tunnel splitting of a macroscopic coherent state.\par
 For particles with a mesoscopic number of spins, the incoherent MQT 
fluctuation rate has the same magnitude as the frequency equivalent of the 
MQC tunnel splitting. Then, instead of trying to detect the MQC state, one
can search for MQT fluctuations with a similar frequency.
In this work, we use $^{57}$Fe M\"ossbauer absorption spectroscopy, 
which has a ``window'' for measurement of electronic fluctuation frequencies
centered at 10$^8$\,Hz, to search for incoherent tunnel
fluctuations in a ferritin sample with a Fe mean loading of 410 atoms per 
core, down to a temperature of 34\,mK. The mean Fe content of our particles is
lower than that of the smallest particles (1000 Fe atoms per core)
where the resonance was found \cite{awschalom,gider}, but the presence 
of a small, but sizeable, unscreened magnetic field in our experiment can be 
thought to preclude the establishment of an eventual MQC state. However, a 
check of the existence
of incoherent fluctuations around a frequency 10$^8$\,Hz in 
ferritin at 34\,mK  can shed light on the existence of a MQC state with
a tunnel splitting of similar frequency at the same low temperature. \par

Details about natural and artificial ferritin can be
found in Ref.\onlinecite{stpier}.
Our artificial ferritin sample was prepared from ferrous Mohr salt with 
iron 95\% enriched in $^{57}$Fe in order to obtain a good signal to noise 
ratio in the M\"ossbauer spectra.
The mean iron content of 410 Fe atoms per core was checked by atomic 
adsorption analysis, and 
the protein concentration in the solution (10\,mg/ml) was determined by the 
Lowry method. The mean distance between particles in the sample is about 
70\,nm. 
The size histogram, established from transmission electron microscopy (TEM)
pictures, is centered at a diameter value $d_0$=4\,nm and has
a lognormal mean square deviation $\sigma$=0.18. The peak in the Zero Field
Cooled (ZFC) branch of the magnetic susceptibility occurs
at $T_m$=7.5\,K. Our particles have very similar mean size and ZFC peak 
temperature as those refered to
as 500 Fe atoms per core in Ref.\onlinecite{gider} ($d_0$=3.9\,nm and 
$T_m$=7.5\,K).\par
The $^{57}$Fe M\"ossbauer absorption spectroscopy experiments were performed 
in zero field at 34\,mK and in the temperature range 4.2\,K -- 40\,K.
The 34\,mK spectrum was
recorded with a spectrometer coupled to a $^3$He--$^4$He dilution refrigerator.
The sample holder is a copper cell with 0.5\,mm thick plexiglass windows, which
can contain about 1\,ml of solution. For the 34\,mK spectrum, the particles
were diluted by a factor 3 to reduce interaction effects.
Thermalisation is achieved by gluing
thin sheets of pure Al (for the spectra at and above 4.2\,K) or of pure
Cu (for the spectra in the dilution refrigerator) onto the plexiglass windows.
The thin metal sheets are in contact with the bulk copper of the sample holder,
thus realising an isothermal box insuring a good thermalisation of the frozen
solution. In the dilution refrigerator, the sample holder is thermally coupled
to the mixing chamber through a cold silver finger.
For the 34\,mK spectrum, the magnetic field at the sample place, created 
mainly by the magnets in the
M\"ossbauer driving unit, was screened, resulting in a maximum field
of 0.15\,G.\par

The magnetic structure of the ferritin cores is expected to be of AF type,
with a N\'eel temperature of a few hundred Kelvins. In the 
simplest picture of N\'eel's model,
an AF lattice consists of two FM sublattices, which have
opposite magnetisations ${\bf M_1}$ and ${\bf M_2}$ with the same magnitude 
$M_0$. The AF ordering can be characterized by the unit N\'eel vector:
\begin{eqnarray}
\bf {l}={{{\bf {M_1-M_2}}} \over {2 M_0}}.
\label{neelvect}
\end{eqnarray} 
At a given temperature $T$, the N\'eel vector fluctuates by crossing 
the energy barrier created by the magnetic anisotropy: this is the 
superparamagnetic relaxation. For an axial anisotropy and for a
given particle volume $V$, the thermally
activated fluctuation frequency for the reversal of the N\'eel vector of a 
particle is given by \cite{neel2,brown}:
\begin{eqnarray}
{1 \over \tau}={ 1 \over \tau_0}\ \exp(-{{KV}\over {k_BT}}),
\label{tauN}
\end{eqnarray}
where K is the magnetic anisotropy energy per unit volume and
$\tau_0$ a microscopic relaxation time with magnitude 10$^{-9}$-10$^{-11}$\,s.
When the thermal energy is much smaller than the barrier height, the N\'eel
vector is classically forbidden from crossing the anisotropy energy barrier, 
but it can tunnel through it. In zero magnetic field and in the limit
of weak dissipation, a macroscopic coherent state can be established where all 
the Fe$^{3+}$ moments in the AF sublattices tunnel between opposite directions 
in perfect unison. The system then presents a tunnel splitting $\Delta$
separating
the antisymmetric and symmetric states built up from the ``up'' and ``down''
configurations of the N\'eel vector {\bf l}.
In the presence of a magnetic field such that the Zeeman splitting is larger
than the tunnel splitting, the quantum coherent state is destroyed, but 
incoherent tunnel fluctuations with frequency $\Gamma$ can take place, with 
reversal of the {\bf l} 
vector (MQT). For a given particle volume $V$, $\Delta$ and $\Gamma$ can be 
expressed, in the WKB approximation, as \cite{gara}:
\begin{eqnarray}
\Delta \simeq \Gamma_0 \exp[-{ {S(V)} \over 2}] \ {\rm and}\ 
\Gamma \simeq \Gamma_0 \exp [-S(V)],
\label{splitting}
\end{eqnarray}
where $\Gamma_0$ is a microscopic frequency and $S(V)$ is the WKB action, 
proportional to $V$. Expressing the action in the AF case yields a  
resonance frequency $\Delta$ in the MQC state \cite{barbara}:
\begin{eqnarray}
\Delta=\Gamma_0 \exp(-{V\over \mu_{B}} \sqrt{\chi_{\bot}K}),
\label{nub}
\end{eqnarray}
where $\chi_{\bot}$ is the AF transverse susceptibility.
The crossover temperature $T^*$ between the thermally activated regime and
the quantum regime, estimated as the temperature where
the rate of the thermal fluctuations is equal to that of the MQT, then writes
\cite{barbara}:

\begin{eqnarray}
T^*={\mu_{B}\over k_B}\sqrt{K/ \chi_{\bot}}.
\label{Tcr}
\end{eqnarray}
Estimations of $T^*$ using standard values for $K$ (10$^6$\,ergs/cm$^3$) and 
$\chi_\bot$ (10$^{-4}$\,emu/cm$^3$) show
that it must be in the range 0.1 -- 1\,K \cite{barbara}.\par

\begin{figure}
\epsfxsize= 4 cm
\centerline{\epsfbox{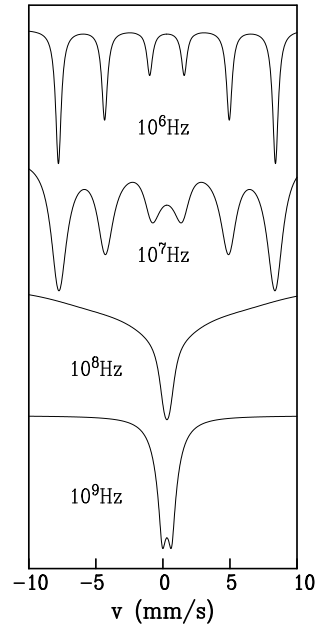}}
\vspace{0.5cm}
\caption{\small\sl $^{57}$Fe M\"ossbauer lineshapes in the presence of
fluctuations of a hyperfine field of 500\,kOe as a function of the relaxation 
frequency, calculated following Ref.\protect\onlinecite{vander}.}
\label{spfl}
\end{figure}

It is well known that the electronic fluctuations frequencies can be measured 
by M\"ossbauer spectroscopy if they lie within a `` relaxation 
window'' centered at the hyperfine Larmor frequency $1/\tau_{\rm L} =
{1 \over {2\pi}}\omega_{hf}$, where $\hbar \omega_{hf}$ is the hyperfine energy
\cite{vander}. For $^{57}$Fe and a magnetic hyperfine field of 500\,kOe,
$1/\tau_{\rm L}$ is
around 10$^8$\,Hz. When the N\'eel vector {\bf l} fluctuates, the hyperfine 
field {\bf H$_{hf}$} at the nucleus, proportional to the individual Fe$^{3+}$ 
magnetic moment (and opposite to it in direction), fluctuates at the same
rate. The M\"ossbauer ``relaxation window'' in the case of a fluctuating
hyperfine field extends over two frequency decades, from 5$\times$10$^6$\,Hz 
to about 2$\times$10$^8$\,Hz; it is illustrated in Fig.\ref{spfl}, which shows 
the evolution of the spectrum as the fluctuation frequency 1/$\tau$ increases.
When 1/$\tau$ is above a few 10$^8$\,Hz, the magnetic hyperfine structure is
smeared out, 1/$\tau$ is no longer measurable, and the spectrum is a single
line or a two-line quadrupolar spectrum. 
However, in the presence of a large distribution of relaxation frequencies, 
as it is
the case for an assembly of superparamagnetic particles, one actually only
observes the patterns corresponding to two populations of particles, those 
whose N\'eel vector fluctuates with a frequency higher than $1/\tau_{\rm L}$ 
(two-line spectrum), and those with 
$1/\tau < 1/\tau_{\rm L}$ (six-line spectrum).\par
\begin{figure}
\epsfxsize= 5cm
\centerline{\epsfbox{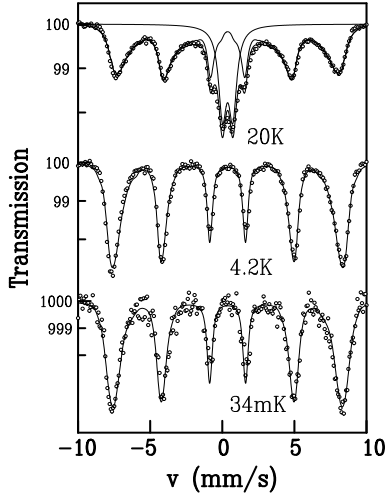}}
\vspace{0.5cm}
\caption{\small\sl $^{57}$Fe M\"ossbauer absorption spectra at selected 
temperatures in zero field in the artificial ferritin sample with a mean 
Fe loading of 410 atoms per core. The solid lines are fits to a distribution
(histogram) of hyperfine fields and, at 20\,K, the fit includes a quadrupolar 
doublet.}
\label{specamb}
\end{figure}

Representative $^{57}$Fe M\"ossbauer absorption spectra in the artificial 
ferritin sample with 410 Fe atoms per core are shown in Fig.\ref{specamb}. 

\begin{figure}
\epsfxsize= 6cm
\centerline{\epsfbox{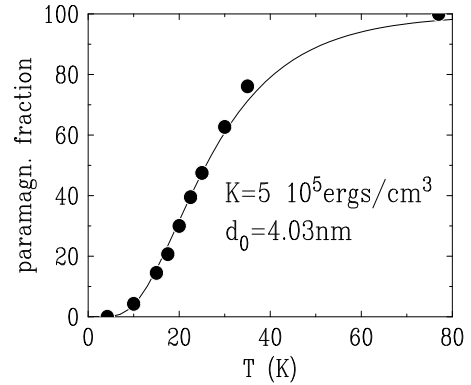}}
\vspace{0.5cm}
\caption{\small\sl Thermal variation of the relative intensity of the 
quadrupolar doublet in the artificial ferritin sample with a mean Fe loading 
of 410 atoms per core. The solid line is a fit using 
Eqn.(\protect\ref{doublet}).}
\label{pourc}
\end{figure}

Above 4.2\,K, the spectra consist of a superposition of a six-line magnetic 
hyperfine field pattern, with a mean hyperfine field $H_{hf} \simeq$490\,kOe,
and of a two-line quadrupolar hyperfine pattern (see spectrum at 20\,K in
Fig.\ref{specamb}). With increasing temperature, the quadrupolar doublet 
progressively replaces the six-line magnetic pattern.
The thermal dependence of the relative intensity $f_p(T)$ of the quadrupolar 
subspectrum is shown in Fig.\ref{pourc}. It can be calculated with the
assumption of thermally activated fluctuations according to N\'eel's
equation (\ref{tauN}). At a given temperature, the distribution of anisotropy 
barriers $KV$ results in a very broad distribution of relaxation times, and
a ``blocking volume'' $V_b(T)$ can be defined such that $\tau = \tau_{\rm L}$:
$V_b(T)={ {k_B T} \over K} \ln(\tau_{\rm L}/\tau_0)$.
Then the ``superparamagnetic fraction'' $f_p(T)$ is given by:

\begin{eqnarray}
f_p(T)={1 \over {\langle V \rangle}} \int_{V_{min}}^{V_b(T)}Vf(V)dV,
\label{doublet}
\end{eqnarray}

\noindent where $f(V)$ is the volume distribution.
The experimentally determined $f_p(T)$ is well reproduced by 
expression (\ref{doublet}), as shown by the solid line in Fig.\ref{pourc}. 
The fit, where the volume distribution function determined from the TEM size 
histograms was used, yields $K=5 \times 10^5$\,ergs/cm$^3$, which is a
standard value for ferric oxides.
Above 4.2\,K, the dynamics of the N\'eel vector
is therefore well described by thermally activated fluctuations across an
anisotropy barrier with mean value: $\langle KV \rangle$=120\,K.
The spectrum at 34\,mK, shown at the bottom of Fig.\ref{specamb}, is 
essentially identical to the spectrum obtained at 4.2\,K. Both spectra can be
fitted with identical narrow distributions of static hyperfine fields.
Such distributions are due to the presence of inhomogeneities
in the lattice of the ferritin cores, and result in a spread of hyperfine field
values of about 10\% on each side of the mean value (500\,kOe). They give rise
to ``inhomogeneous'' static broadenings of the M\"ossbauer lines, which remain
unchanged below 4.2\,K. 
Therefore, at 34\,mK, all the particles have 
their N\'eel vector fluctuating with frequencies below the ``relaxation 
window'', i.e. slower than 5$\times$10$^6$Hz.\par

Several solutions of artificial ferritin, 
with an iron average loading ranging from 1000 to 4000 Fe atoms per core, were
studied in Ref.\onlinecite{awschalom,gider}. 
In these studies, the chemical process used to synthesize the samples was 
identical to ours.
For all the samples, a well-defined resonance was observed below 200\,mK, 
both in the magnetic noise $S(\omega$) and in the imaginary part 
$\chi''(\omega)$ of the a.c.
susceptibility, which was attributed to the tunnel splitting associated
with MQC. The obtained resonance frequency increases rapidly as the particle
size decreases and, for the samples with a Fe loading less than
2000 atoms, the resonance frequency at 25\,mK is found to be at 10$^8$\,Hz or
higher, which is precisely of the same magnitude as the hyperfine Larmor 
frequency associated with $^{57}$Fe M\"ossbauer spectroscopy. 
Therefore, if a macroscopic coherent state could
establish in the 25 -- 50\,mK range in the ferritin samples with the smallest 
Fe loading, with a tunnel splitting around or above 10$^8$\,Hz, then 
incoherent tunnel fluctuations with a similar frequency should be observed. 
Their signature in the M\"ossbauer spectra would be a single line 
or a two-line quadrupolar pattern (see
the simulated spectra at 10$^8$ and 10$^9$\,Hz in Fig.\ref{spfl}), and such
a signature is absent from the 34\,mK spectrum. Therefore, there are no 
incoherent tunnel
fluctuations of the N\'eel vector at 34\,mK with a frequency around 10$^8$Hz
in our artificial ferritin sample with 410 Fe atoms per core.\par

Our $^{57}$Fe M\"ossbauer experiments were however not performed exactly
in the same conditions as the $\chi(\omega)$ experiments of 
Ref.\onlinecite{gider,awschalom}.
First, the magnetic field at the sample location is larger in
our experiment (15\,$\mu$T $vs$ 1\,nT). Such a field creates a Zeeman
splitting between the ``up'' and ``down'' configurations of the N\'eel
vector of the order of 10$^6$\,Hz per percent of uncompensated moment; this
splitting is smaller than the supposed tunnel splitting of 10$^8$\,Hz
and is neither expected to destroy the MQC state nor to have an influence on 
existing incoherent tunnel fluctuations. Second, our ferritin sample has been 
prepared with Fe 95\% enriched 
in the isotope $^{57}$Fe, which has a ground nuclear spin I=1/2, and thus
a non-zero nuclear magnetic moment, whereas the samples used in 
Ref.\onlinecite{gider,awschalom}
were made from natural Fe, where the abundance of $^{57}$Fe is 2.2\%. 
Actually, all the nuclear spins in the sample (i.e. both the $^{57}$Fe and 
proton spins) can be expected to have an influence on the tunneling properties.
In this respect, there is no great difference as to the total nuclear spin 
content between our $^{57}$Fe enriched sample and those made with natural Fe. 
It is generally admitted that a MQC state is much
more likely to be destroyed by the coupling to the nuclear spins (both through
dissipation \cite{garg1} and decoherence effects \cite{stamp}) than MQT. 
Concerning MQT fluctuations, recent models show that the nuclear spins 
provide a ``spin bath'' which is likely to enhance the tunnel relaxation rate
\cite{proko,mqt} rather than deplete it. \par

Ideally, the check for the existence of tunnel fluctuations by M\"ossbauer 
spectroscopy should have been performed in the same conditions as the 
experiments of Ref.\onlinecite{awschalom,gider}, i.e. in a ferritin sample 
prepared with natural Fe. Unfortunately, such an experiment with small Fe 
loadings would
require impractically long counting times because of the very small signal.
However, the presence of nuclear moments is not expected to influence at all
the thermally activated fluctuations of the N\'eel vector. Then, at least
down to 4.2\,K, the spectra of an artificial ferritin sample with a natural Fe
loading of 410 atoms per core would be identical with that of the same sample
prepared with $^{57}$Fe, as the crossover temperature to the quantum
regime is expected to lie below 4.2\,K  for
AF particle systems \cite{barbara}. As the fluctuation frequencies at 
4.2\,K in our
sample are below the M\"ossbauer ``relaxation window'' (i.e. they are lower
than 5$\times$10$^6$\,Hz), then a spectrum at 34\,mK in a
ferritin sample with natural Fe loading would (if feasible) also show that
all the fluctuation frequencies are below 5$\times$10$^6$\,Hz. Therefore,
one can conclude that incoherent tunnel fluctuations at a frequency 
10$^8$\,Hz are absent at 34\,mK in artificial ferritin with a mean Fe loading 
of 410 atoms per core, implying the absence of MQC of the N\'eel vector at
that temperature.\par

As a conclusion, our $^{57}$Fe M\"ossbauer experiments in artificial ferritin
with a mean Fe loading of 410 atoms per core down to 34\,mK show that, if a 
MQC state can 
establish in ferritin cores with low Fe loading, its tunnel splitting
cannot be in the 10$^8$\,Hz range. This casts a doubt on the interpretation,
in similar artificial ferritin samples, of a
resonance at a frequency 10$^8$\,Hz as due to a tunnel splitting in a
macroscopic coherent state \cite{gider,awschalom}. Indeed, one then would 
expect 
incoherent tunnel 
fluctuations to be detectable at 10$^8$\,Hz or above, which we do not
observe.
Our data however do not exclude 
tunnel fluctuations frequencies slower than $5 \times 10^6$\,Hz, as these 
would not be observable by $^{57}$Fe M\"ossbauer spectroscopy.

\end{multicols}

\end{document}